\documentclass[journal]{journal}
\usepackage{graphicx}
\usepackage{amssymb}
\usepackage{amsmath}

\begin{document}

\title{Socializing Autonomous Units with the Reflexive Game Theory and Resonate-and-Fire neurons}


\author{Sergey~Tarasenko
\thanks{S. Tarasenko is an independent researcher.  Email: infra.core@gmail.com
}
}



\maketitle

\begin{abstract}
In this study the concept of reflexia is applied to modeling behavior of autonomous units. The relationship between reflexia, on the one hand, and mirror neuron system and perception of emotions, on the other hand, is introduced. The main method of using reflexia in a group of autonomous units is Reflexive Game Theory (RGT). To embody RGT in a group of autonomous agents a communication system is employed. This communication system uses frequency domain multiplexing by means of Izhikevich's resonate-and-fire neural models. The result of socialization of autonomous units by means of RGT and communication system is illustrated in several examples. 
\end{abstract}


\section{Introduction}
\label{intro}
The notion of $reflexia$ in the psychological context was fisrt introduced by Lefebvre in late 60s \cite{lef3,lef7,lef8}.  

$Reflexia$ means projection of the external world on one's mental state. More specifically, if a human being stands 
in the field of barley he/she can $imagine$ onesself standing in the field of barley from the 3rd person's perspective. Thus, preserving the egocentric point of view humans are capable of imaganing their own allocentric representation. \footnote{Term $egocentric$ means perception from the 1st person perspective, while term $allocentric$ means the perception from the 3rd person perspective.} 

Therefore, the gist of reflexia is an ability to imagine self perception in the \textit{allocentric reference frame} (external\footnote{$Extenal$ means here outside of ones body or any other feasable entity} point of view) being operating in the one's \textit{egocentric reference  frame}.

Reflexia is an ability to penetrate into the deeper layers of one's psychological state. An abstract example of penetration into the deeper layers is when subject $a$ can imagine another person (subject $b$), who is imaganing subject $a$, the world around and himself imagaging it.

The overall results about the justification of reflexia and its application for modeling human behavoir has been summarized by Vladimir Lefebvre, the principle investigator in this field, in his book ``Algebra of Consciece" \cite{lef5,lef6}.

Here we would like to refer two more examples of potential implementation of reflexia: \textit{mirror neuron system} and \textit{perception of emotions}. 

\textit{Mirror neuron system}: The key concept of the Mirror Neuron System discovered by Rizzolatti and his colleagues \cite{rizzo,rizzoetal} is that there are neurons in primate brain that activate in both cases when primate is doing a particular action itselt or observe someone else doing the same action. Therefore mirror neuron system translates external state of another agent into the internal state of the current agent. Therefore primates can repeat the observed action. This functionallity is very close to the notion of $reflexia$.

\textit{Perception of emotions}: Usually, the emotions are characterized by some physiological patterns of body activity on the one hand, and external expression by face mimic or gestures on the other hand. A reproduction by the onesself of physical part of emotional pattern, i.e., just making an angry face, can elicit the anger as emotional state itself \cite{ekman,leven}. This suggests that there is a mechanism that helps mapping someones internal state to the self.  For instance, Edgar Allan Poe, in his story ``The Purloined Letter'', describes how one character is attemping to understand the intensity of emotional experience of another character by self-mimicing (or imitating) facial expressions of another one. 

Most recently the principle of reflexia have been reconsidered in the shape of the Reflexive Game Theory (RGT). The expectations are that since reflexia is intrinsic ability of human being and the principles and models proposed by Lefebvre have been proved to be true, the Reflexive Game Theory can deliver the human-like decision-making processes.

In contrast to the Game Theory based on purely utilitary and rational principle, the Reflexive Game Theory is based on the human's decision making. The conner stone of the Reflexive Game Theory is the \textit{egoism forbiddeness principle}, while Game Theory is build upon pure egoism stemming from (Min Loss - Max Utility) principle.

Most recently RGT has been applied for theoretical modeling of human-robot groups \cite{taras}, in which robots were sucessfuly refraining people from doing risky actions. However, the issue of how to embody RGT into the system of robotic agents is still an open questions.

The goal of this paper is to illustrate how application of RGT algorithms for modeling purely robotic groups can literary humanize the robots by given them human sensitivity instead of cold rationality, which is usually attributed to machines. A certian structure of communication system to enable robots ``talk" to each other is proposed.

\section{Brief Overview of the Reflexive Game Theory}
\label{overview}

\subsection{Representation of groups: graphs, polynomials, stratification tree and decision equation}
\label{repres}

The exhaustive desription of the Reflexive Game Theory (RGT) and tutorial of RGT application have been presented by Lefebvre \cite{lef1,lef2,lef4}. Here, we present a brief overview of RGT enough to understand its basic concept and formulate the tasks solved in this paper. 

The RGT deals with groups of abstract subjects (individuals, humans, autonomous agents etc). 
Each subject is assigned a unique variable (\textit{subject variable}). Any group of 
subjects is represented in the shape of \textit{fully connected graph}, which is called 
a \textit{relationship graph}. Each vertex of the graph corresponds to a single subject.The name of each vertex is a unique subject variable.

The RGT uses the set theory and the Boolean algebra as the basis for calculus. Therefore the values of subject variables are elements of Booleans algebra. 

All the subjects in the group can have either alliance or conflict relationship. The relationships are illustrated with graph ribs. The solid-line ribs correspond to alliance, while dashed ones are considered as conflict. For mathematical analysis alliance is considered to be conjunction (multiplication) operation ($\cdot$),  and conflict is defined as disjunction (summation) operation (+). 

\begin{figure}
\centering
\includegraphics[height=2cm]{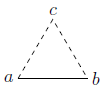}
\caption{Relationship graph.}
\label{relGrp}       
\end{figure}

The decomposable relationship graphs \cite{lef1,lef2,batlef} can be presented in the analytical form of a corresponding \textit{polynomial}. Any relationship graph of three subjects is decomposable. Consider three subjects $a, b$ and $c$. Let subject $a$ is in alliance with other subjects, while subjects $b$ and $c$ are in conflict (Fig.~\ref{relGrp}). The polynomial corresponding to this graph is $ab+c$.

\begin{figure}
\centering
\includegraphics[height=2cm]{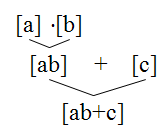}
\caption{Polynomial Stratification Tree. Polynomials $[a], [b]$ and $[c]$ 
are elementary polynomials.}
\label{fig:fig2}
\end{figure}

Regarding a certain relationship, the polynomial can be stratified (decomposed) into \textit{sub-polynomials} 
\cite{lef1,lef2}. Each sub-polynomial belongs to a particular level of stratification. If the stratification regarding conflict (alliance) was first built, then the stratification regarding alliance (conflict) is implemented on the next step The stratification procedure finalizes, when the \textit{elementary polynomials}, containing a single variable, for each variable are obtained after a certain stratification step.
The result of stratification is the \textit{Polynomial Stratification Tree (PST)}. It has been proved that each non-elementary polynomial can be stratified in an unique way, i.e., each non-elementary polynomial has only one corresponding PST (see \cite{batlef} considering one-to-one correspondence between graphs and polynomials). Each higher level of the tree contains polynomials simpler than the ones on the lower level. For the purpose of stratification the polynomials are written in square brackets. The PST for polynomial $ab+c$ is presented in Fig.\ref{fig:fig2}.

We omit the branches of the PST and from each non-elementary polynomial write in top right corner its sub-polinomials. The resulting tree-like structure is called a \textit{diagonal form}\cite{lef1,lef2,lef5,lef6}. Consider the diagonal form corresponding to the PST presented in Fig.~\ref{fig:fig2}:
\[\begin{array}{*{20}{c}}
   {} & {} & {[a][b]} & {}  \\
   {} & {[ab]} & {} & { + [c]}  \\
   {[ab + c]} & {} & {} & {}  \\
\end{array}\]

We introduce the \textit{universal set}, which contains the \textit{elementary actions}. For example, these actions are actions $\alpha$ and $\beta$. The \textit{Boolean algebra} of the universal set includes four elements: $1 = \{\alpha , \beta \}$, $ \{\alpha\}$, $\{\beta\}$ and the empty set 0 = $\{\}$ = \O. The diagonal form is considered to be a function defined on the Boolean algebra. 

Accroding to definition given by Lefebvre \cite{lef5}, we present here exponential operation defined by formula 
\begin{equation}
\label{expfrom}
P^W = P + \overline{W} \ , 
\end{equation}

where $\overline{W}$  stands for negation of $W$ \cite{lef1,lef2,lef4}. It can be shown that function in eq. (\ref{expfrom}) is equivalent to \textit{implication function} \cite{lef6}.

This operation is used to fold the diagonal form. During the folding, round and square brackets are considered to be interchangeable. Next we implement folding of diagonal form of polynomial $ab+c$:
\[\begin{array}{*{20}{l}}
   {} & {} & {[a][b]}& {} & {}   \\
   {} & {[a][b]} & {} &{+[c]}  \\
   {[ab + c]} & {} & {} & {}& { = ab + c} 
\end{array}\]

\subsection{The Decision Equation: definition and solution}
\label{deceq}

Each subject in the group should choose an alternative (element) of the Boolean algebra. We consider the \textit{decision equations}. Each equation corresponds to a single subject. The solution of such equation defines the choice of each subject in the group. This equation contains subject variable in the left-hand side and the result of diagonal form folding in the right-hand side: 
\begin{eqnarray}
a = ab + c  \\
b = ab + c  \\
c = ab + c
\end{eqnarray}
To solve the decision equation, one should first transform it into canonical form \cite{lef1,taras} defined as:
\begin{equation}
\label{canfrom}
x = Ax + B\overline{x}  \;  ,
\end{equation}
where  $x$  is the subject variable, and  $A$ and $B$ are some sets.  

This equation has solution if and only if the set $B$ is contained in set $A$: $A \supseteq B$. If this requirement is satisfied, then eq.(\ref{canfrom}) has at least one solution from the interval $A \supseteq x \supseteq B$ \cite{lef1,lef2,lef4}. Otherwise, the decision equation has no solution, and it is considered that subject cannot make a decision. The state of inability to make decision (choice) is called a $frustration$. The explicit  tranformation of decision equation for subjects $a$, $b$ and $c$  into canonical form is consider in \cite{taras2}. Therefore, here we only provide the resulting canonical forms:
\begin{eqnarray}
a = (b+c)a + c\overline{a} \label{cfde1} \\
b = (a+c)b + c\overline{b} \label{cfde2}\\
c = c + ab\overline{c} \label{cfde3}
\end{eqnarray}

Next we consider two tasks, which can be formulated regarding the decision equation in the canonical form and briefly discuss methods to solve each task.

\subsection{The Forward and Inverse Tasks}
\label{forwardT}

In this section, we only illustrate the introductory examples of the Forward and Inverse task of the RGT. The comprehensive explaination how to solve the forward task can be found here \cite{lef1,lef2,lef4}. The issues regarding the Inverse task are discussed in details in \cite{taras2}.

\textit{The Forward Task}. The variable in the left-hand side of the decision equation in canonical form is the variable of the equations, while other variables are considered as influences on the subject from the other subjects. The  \textit{forward task} is formulated as a task to find the possible choices of a subject of interest, when the influences on him from other subjects are given. 

\begin{table}
\caption{Influence Matrix}
\label{inflmatrix}
\begin{center}
\begin{tabular}{|c|c|c|c|}
\hline
{}&a&b&c\\
\hline
\rule{0pt}{12pt}a&a&$\{\alpha\}$&$\{\beta\}$\\[2pt]
\hline
\rule{0pt}{12pt}b&$\{\beta\}$&b&$\{\beta\}$\\[2pt]
\hline
\rule{0pt}{12pt}c&$\{\beta\}$&$\{\beta\}$&c\\[2pt]
\hline
\end{tabular}
\end{center}
\end{table}

The mutual influences in the forward task are presented in the \textit{Influence matrix} (Table \ref{inflmatrix}).
The main diagonal of influence matrix contains the subject variables. The rows of the matrix represent influences of the given subject on other subjects, while columns represent the influences of other subjects on the given one. The influence values are used in decision equations.

I illustrate solution of the forward task using subjects $a$, $b$ and $c$. By using canonical forms of decision equations for each subject (eqs. \ref{cfde1}-\ref{cfde3}) and Influence matrix, we obtain the choice of each subject:

subject $a$: $a = (\{\beta\}+\{\beta\}) a + \overline{a}  \Rightarrow  a = \{\beta\}a + \overline{a}$.

subject $b$: $b = b + (\{\alpha\}\{\beta\}+\overline{\{\alpha\}})\overline{b}    \Rightarrow b = b + 
\{\beta\}\overline{b}$.

subject $c$: $c = c + (\{\beta\}\{\beta\}+\overline{\{\beta\}})\overline{c}  \Rightarrow 
c = c+(\{\beta\} +\{\alpha\})\overline{c}  \Rightarrow  c = 1$.

Equation for subject $a$ does not have any solutions, since set $A = \{\beta\}$ is contained in set $B = 1$: $A \subset B$.  Therefore, subject $a$ cannot make any decision. Therefore he is considered to be in frustration state.

Equation for subject $b$ has at least one solution, since $ A = 1 = \{\alpha, \beta\}\supseteq B = \{\beta\}$. The solution belongs to the interval $1\supseteq b \supseteq \{\beta\}$. Therefore subject $b$ can choose any alternative from Boolean algebra, which contains alternative $\{\beta\}$. These alternatives are $1 = \{\alpha,\beta\}$ and $\{\beta\}$.

Equation for subject $c$ turns into equality $c = 1$. This is possible only in the case, when $A = B = 1$. 

The solution of the Forward task can be algorithmized as follows:

1) formalize of actions in terms of Boolean algebra of alternatives;

2) represent a group as relationship graph;

3) represent relationship graph 

\textit{The Inverse Task}. The \textit{inverse task} is formulated as a task to find all the simultaneous (or joint) influences of all the subjects together on the subject of interest that result in choice of a particular alternative or subset of alternatives. We call the subject of interest to be a \textit{controlled subject}.

Let subject $a$ is the controlled subject and $a^*$ is a fixed value representing an alternative or subset of alternatives, which subjects $b$ and $c$ want subject $a$ to choose. By substituting subject variable $a$ in decision equation for fixed value $a^*$, we obtain the \textit{influence equation}:
\begin{equation}
a^* = (b+c)a^* + c\overline{a^*}  \;  ,
\end{equation}

In contrast to the decision equation, which is equation of a single variable, the influence equation is the equation of multiple variables.  We need to find all the joint influences of subjects $b$ and $c$ in form of pairs $(b,c)$. Let $a^* = \{\alpha\}$, then we need to solve the system of equations

\begin{equation}
  \begin{cases}
  b + c = \{\alpha\} \\ 
  c = \{\alpha\}  
  \end{cases}
\label{sys1}
\end{equation}

Consequently, we have to solve equation 
\begin{equation}
b + \{\alpha\} = \{\alpha\}  
\label{eq1sys1}
\end{equation}
and to find all the pairs $(b,c)$, results in solution  $a^* = \{\alpha\}$. These pairs are solutions of the system (\ref{sys1}). Therefore, we run all the possible values of variable $b$ and check if the first equation of the system (\ref{eq1sys1}) turns into true equality: 

$b = 1: 1 + \{\alpha\} = 1 \Rightarrow 1 \neq \{\alpha\}$;

$b = \{\alpha\}: \{\alpha\} + \{\alpha\} = \{\alpha\} \Rightarrow \{\alpha\} = \{\alpha\}$

$b = \{\beta\}: \{\beta\} + \{\alpha\} = 1 \Rightarrow 1 \neq \{\alpha\}$; 

$b = 0: 0 + \{\alpha\} = \{\alpha\} \Rightarrow \{\alpha\} = \{\alpha\}$. 

Therefore, out of four possible values, only two values $\{\alpha\}$ and $0$ are appropriate. Thus, we obtain two pairs $(b,c)$: $(\{\alpha\},\{\alpha\})$ and $(\{\alpha\},0)$. These pairs represent the strategies of \textit{reflexive control}.

\begin{figure}
\centering
\includegraphics[width=8.5cm]{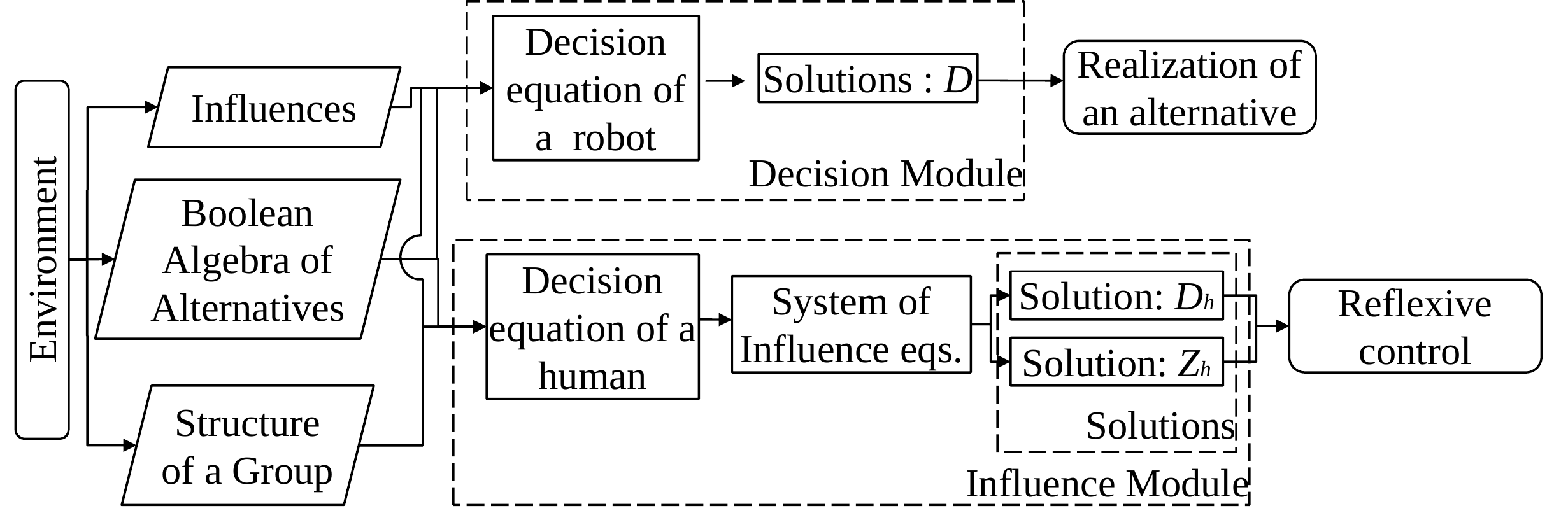}
\caption{The Basic Control Schema of an Abstract Subject (BSCAS).}
\label{bcs1}
\end{figure}

This conludes the overview of the Reflexive Game Theory. As a final remark, we show Basic Control Schema of Abstract Individual (BSCAS) (Fig.~\ref{bcs1}). The detailed decription of the BSCAS can be found in \cite{taras2}.

Summarizing this section, I emphasize the information needed for the RGT to be applied. First of all, we need to define the universal set of actions and the corresponding Boolean algebra. Second, we need to know the structure of a group. Finally, we need to know the mutual influences of the group members. 

This imposes the following requirements of functionallity of autonomous units. Autonomous units have to be able to 1) code each alternative in the Boolean algebra and relationships; 2) trasmit this information to each other in a manner that each unit could `understand' the information address to it and at the same time all the units should be aware about all the information transfered from any unit to any unit.

How to code and transfer this information in the groups of autonomous units is the main question discussed in the rest of this study. Therefore, the material presented hereafter is dedicated to the matter of how an autonomous unit can obtain the required information.




\section{Introducing Communications between Autonomous Units}
\label{mobunits}

In this study, we use the abstract autonomous units, which are capable of communicating with each other in the frequency domain. Therefore, these units can distinguish between several frequencies. Each frequency can be used as a carring frequency to transmit the information between the autonomous units.

To be capable of distinguishing between various frequencies the autonomous units are supplied with frequency selective devices - resonators. There are many possible implimentations of the resonators. In this study, we suggest to used the \textit{Resonate-and-Fire} linear neural model proposed by Izhikevich \cite{izhikev}. The choice of this model is justified by its simplicity in implementation and low computational complexity. Next, we provide brief overview of the \textit{Resonate-and-Fire} linear model. 

\subsection{Resonate-and-fire neurons}
\label{resfire}

The original linear model of resonate-and-fire neuron proposed by Izhikevich \cite{izhikev} is described by 
the system of two \textit{differential equations}:
\begin{equation}
  \begin{cases}
     \dot{x} = bx - \omega y \\
     \dot{y} = \omega x + by  
  \end{cases}
\label{resfire}
\end{equation}
where $x$ is current-like or recovery vairable, and $y$ is voltage-like or action potential variable, in terms of neuroscience. Both variables $x$ and $y$ are functions of time: $x = x(t)$ and $y = y(t)$. Notation $\dot{x}$ means derivative of $x$ with respect to time $t$: $\dot{x} = \partial x / \partial t$. Paramater $\omega$ is the eigen-frequency, which is preferred or resonant frequency of the system, and represents frequency of subthreshold oscillations; parameter $b$ is ananalog of damping factor in the linear damped oscillator. The value of paramter $b$ is set to -0.1 throughout our simulations.
 
For the purpose of numerical intergration the system (\ref{resfire}) can be transformed into the form:
\begin{equation}
\label{complexResF}
\dot{z} = (b + i\omega)z \ .
\end{equation}

Then variables $x$ and $y$ are real and imaginary parts of complex variable $z$, respectively. Using eq.(\ref{complexResF}) and first-order \textit{Euler method}, we striaghtforwardly obtain \textit{difference equation} (\ref{differ}) from differential equation (\ref{complexResF}):
\begin{equation}
\label{differ}
z(t+\tau) = z(t) + \tau (b + i\omega)z(t),
\end{equation}
where $\tau$ is a small time step. We set $\tau$ to 0.005 in all our simulations. 

Iterating difference equation (\ref{differ}) with $z(0) = z_0$, we can approximate the analitical solution of eq.(\ref{complexResF}) and, consequently, of system (\ref{resfire}).

Now to obtain value of voltage variable $y(t)$ at time $t$, we only need to take imaginary part of $z(k)$.

The real nearons produce a spike, once value of $y(t)$ equal to or exceeds some preset threshold. However, this feature is not provided by the linear model. 

Therefore we slightly modify the original model by adding the `spiking' condition:

\begin{equation}
  \begin{cases}
    if \ \ y(t) \geq threshold \ \  y(t) = 1.5\\
    y(t+\tau) = 0.1
  \end{cases}
\end{equation}

A threshold value is to 1 throughout the simulations. Therefore, the ultimate model of resonate-and-fire neuron used in this study is described by the system (\ref{resfireext}):

\begin{equation}
  \begin{cases}
     z(t+\tau) = z(t) + \tau (b + i\omega)z(t) \\
     if \ \ y(t) \geq 1 \ \  y(t) = 1.5 \\ 
     y(t+\tau) = 0.1 
  \end{cases}
\label{resfireext}
\end{equation}

\begin{figure}
\includegraphics[height=6.8cm]{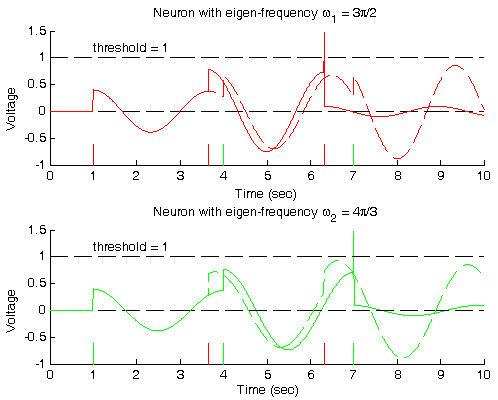}
\caption{The Resonate-and-Fire neurons. $Top$: solid line illustrates resonanse with the input frequency $\omega_1$ = 3$\pi$/2, dashed line shows only subthreshold oscillations meaning that neuron does not respond to the frequency $\omega_2$ = 4$\pi$/3. $Bottom$: solid line illustrates resonanse with the input frequency $\omega_2$ = 4$\pi$/3, dashed line shows only subthreshold oscillations meaning that neuron does not respond to the frequency $\omega_1$ = 3$\pi$/2. The green and red vertical lines indicate the equal input pulses of magnitude 0.4. Green and red pulses are provided with frequencies $\omega_1$ = 3$\pi$/2 and $\omega_2$ = 4$\pi$/3, respectively. Each series of pulses starts 1 ms after the system onset. Threshold is set to 1. Parameter $b$ is -0.1.}
\label{resfireplus}       
\end{figure}

We present the sample dynamics of two resonate-and-fire neurons, described by system (\ref{resfireext}), with different eigen-frequencies $\omega_1$ = 3$\pi$/2 and $\omega_2$ = 4$\pi$/3 in Fig.~\ref{resfireplus}. 

It is illustrated that neurons with eigen-frequency $\omega_1$ spikes for the series of pulses with the same frequency and does not respond to the series of pulses with frequency $\omega_2$. The same is true for the second neurons regarding shift in roles of frequencies $\omega_1$ and $\omega_2$.

Thus, we have described the mechanism of frequency selectivity. This can be used to enable multiple neurons to talk to each other via the same medium by means of \textit{Frequency Domain Multiplexing}.

\begin{figure}
\includegraphics[height=6.8cm]{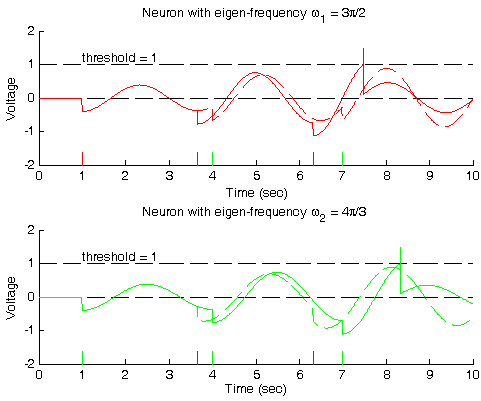}
\caption{The selective responses of resonate-and-fire neurons to the series of equal inhobitory pulses (magnatude -0.4).}
\label{resInh}       
\end{figure}

However, the linear model has other important properties. The inhibitory pulses can also make resonate-and-fire neurons to spike, if the inhibitory pulses are applied with the eigen-frequency of the neuron (Fig.~\ref{resInh}).

\begin{figure}
\includegraphics[height=6.8cm]{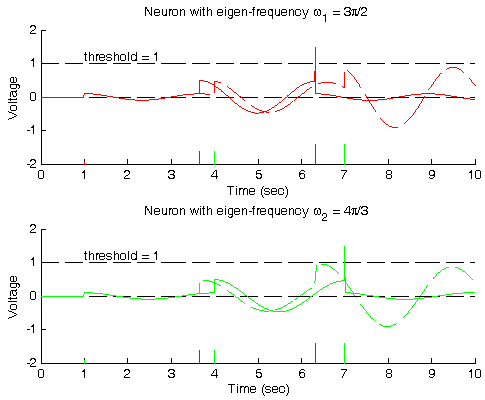}
\caption{The selective responses of resonate-and-fire neurons to the excitatory series \{0.1, 0.4, 0.6\}.}
\label{resmodex}       
\end{figure}

\begin{figure}
\includegraphics[height=6.8cm]{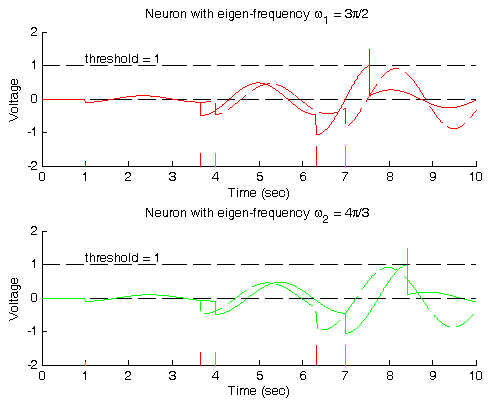}
\caption{The selective responses of resonate-and-fire neurons to the inhobitory series \{-0.1, -0.4, -0.6\}.}
\label{resmodinh}       
\end{figure}

However, it is not the final feature of this model. It is possible to make the neuron fire with series of pulses of different magnitudes. For example, let the magnitudes of the first, second and third pulses are 0.1, 0.4 and 0.6, respectively (Fig.~\ref{resmodex}). The same result will occur for the inhibitory pulses (Fig.~\ref{resmodinh}).

Since the neurons are selective to a certian frequency, it is possible to transfer signals of several frequencies through the same communication channel. 


This concludes the description of communication system. Next we consider the framework to manage the groups of autonomous units.

\section{Building the Groups of Autonomous Units}
\label{groups}

In this section we introduce a sketch of communication system to socialize autonomous units.

\subsection{Information Coding}
\label{unitcom}

Each autonomous unit has several resonators tuned to particular frequencies. Each resonator corresponds to a certian unit in the group. Therefore, the total number of resonators equals the total number of units in the group. For each unit, we reserve its unique frequency. Once the information is obtained from resonators with a frequency accosiated with this unit, it is that this unit is an addressee.

Using the resonate-and-fire neurons presented in the previous section, we can transfer different types of information
throught the network of autonomous units. In fact, we can transmit two types of information coded by 1) the kind of pulses (exictatory vs inhibitory), and 2) selecting different magnitude of pulses in the series.

These two types of information are enough to model the groups in the Reflexive Game Theory. We consider a certain frequency to be the unique identifier of the autonomous unit in the group. Next, if the series pulses contains the excitatory impulses, it is considered that two units are in alliance relationships, or they are in conflict otherwise. Finally, we can define a certian alternative of the Boolean Algebra by a certain seris of pulses.

\subsection{Receiving Informaiton in the Group}
\label{unitcom}

\begin{figure}
\includegraphics[height=6.7cm]{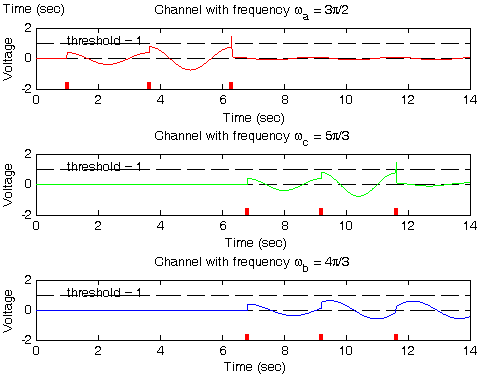}
\caption{Receiving messages in the network.}
\label{installCom}       
\end{figure}

So far, we understand how a certian unit can recieve information. The question remains how autonomous unit can understand where the signal comes from or which unit sends it?

For this purpose, we reserve a series of equal pulses \{0.4,0.4,0.4\}. Hereafter, we refer to the series of pulses as $code$ or $message$. In particular, we call the code \{0.4,0.4,0.4\}  to be \textit{identification-code (ID-code)}, if this code is transmitted by a certain unit on its own preferred frequency. 

Suggest, we have three units $a$, $b$ and $c$. Each unit is characterized by its preferred frequency: $\omega_a = 3\pi/2$, $\omega_b = 4\pi/3$ and $\omega_c = 5\pi/3$.

If autonomous unit $a$ with eigen-frequency $\omega_a$ decides to send some code to another unit, it first sends  \textit{ID-code} on the frequency $\omega_a$. Therefore, the corresponding neuron spikes in each autonomous units, and units $b$ and $c$ 'understand' that unit $a$ wants to send a code. This can be considered as unit $a$ attracts attention of the other units in the group. Then, after a short delay (0.5 sec) after spike on the frequency $\omega_a$, unit $a$ sends a certain code on the frequency $\omega_{subject}$, where $subject$ can be either $b$ or $c$.

As an example, we consider that unit $a$ wants to send its ID-code to unit $c$. Therefore, unit $a$ first sends ID-code \{0.4,0.4,0.4\} on the frequency $\omega_a$ to attract attention of other units: in units $b$ and $c$ the channels with frequency $\omega_a$ show a spike (Fig.~\ref{installCom}, top). Then, 0.5 sec after a spike on the frequency $\omega_a$ , unit $a$ sends the ID-code on the frequency $\omega_c$:  in units $b$ and $c$ the channels with frequency $\omega_c$ show a spike (Fig.~\ref{installCom}, center). Since, frequency $\omega_c$ is the frequency reserved for unit $c$, unit $c$ receives ID-code from unit $a$. At the same time channel with frequency $\omega_b$ shows no spike (Fig.~\ref{installCom}, bottom), and unit $b$ `understands' that ID-code is not addressed to it.

This way each unit in the groups can become completely awear about the whole information transmitted between any two units. Therefore, such communication schema provides all necessary information for application of the Reflexive Game Theory.

\subsection{How to Install Relationships in the Group}
\label{relat}

Now we consider how to install relationship between units. Each unit decides on its own, which type of relationship (conflict or alliance), it wants to install with other units. For example, we consider that the relationships are decided at random, meaning that at the very begining the units do not have any infomation about each other, except for the preferred frequencies. Therefore, this condition can be assumes as guessing. The human guessing based on no prior information has been describe from both theoretical and experimental points of views in \cite{lef6,tarin}. In the case of two options, one option (\textit{positive pole}) is chosen with probability $p \approx 0.61$, while another option (\textit{negative pole}) is chosen with probability $1-p \approx 0.39$. The concept of option's polarity has been first introduced by Lefebvre \cite{lef6}. We consider alliance relationship to be positive pole and conflict relationship to be a negative pole.

The alliance and conflict relationships are coded with codes \{0.4,0.4,0.4\} and  \{-0.4,-0.4,-0.4\}, respectively. We call codes \{0.4,0.4,0.4\} and  \{-0.4,-0.4,-0.4\} to be $alliance$ and $conflict$ codes, respectively, if they are transmitted NOT on the preferred frequency of the unit, which sends it. The alliance and conflict codes are chosen with probabilities 0.61 and 0.39, respectively.

However, to install the relationship, the decisions of both units are needed. In other words, since units chosen the type of relationships independently from each other, it is possible that, for example, unit $a$ sends conflict code to unit $b$, but unit $b$ sends alliance code to unit $a$. Therefore, each unit has decided its own relationship, which is different from the one chosen by counterparty. Thus, the codes are different. We define that the alliance relationship is installed if and only if both units send alliance code to each other, the conflict relationship is installed otherwise. Thus, the relationship between units $a$ and $b$ is conflict. 

If we consider codes \{0.4,0.4,0.4\} and \{-0.4,-0.4,-0.4\} as logic 1 and 0, respectively, the alliance relationship can be defined as logic conjunction (AND) function, and conflict relationships as disjunction (OR) function.

\textit{Example 1.} Let us generate a group of three units with randomly chosen relationships. We use a uniform random number generator with interval (0,1). If the value of random variable exceeds 0.61, unit $x$ generates conflict code \{-0.4,-0.4,-0.4\} to some other unit, otherwise it generates alliance code \{0.4,0.4,0.4\} (Table \ref{relIns}). The rows of Table \ref{relIns} contain the decisions about relationships that each unit generated itself, but have not yet transmitted to other units. Therefore, Table \ref{relIns} illustrates \textit{internal state} of each unit. This internal state is yet not known by other units in the group.

\begin{table}
\caption{Internal state of each unit regarding relationships with others}
\begin{center}
\label{relIns}
\begin{tabular}{|c|c|c|c|}
\hline
{}&a&b&c\\
\hline
\rule{0pt}{12pt}a& - &$0.81$& 0.92\\[2pt]
\hline
\rule{0pt}{12pt}b&0.63&-&0.12\\[2pt]
\hline
\rule{0pt}{12pt}c&0.09&0.27&-\\[2pt]
\hline
\end{tabular}
\end{center}
\label{table:tab1}
\end{table}

\begin{figure*}
\includegraphics[height=7cm]{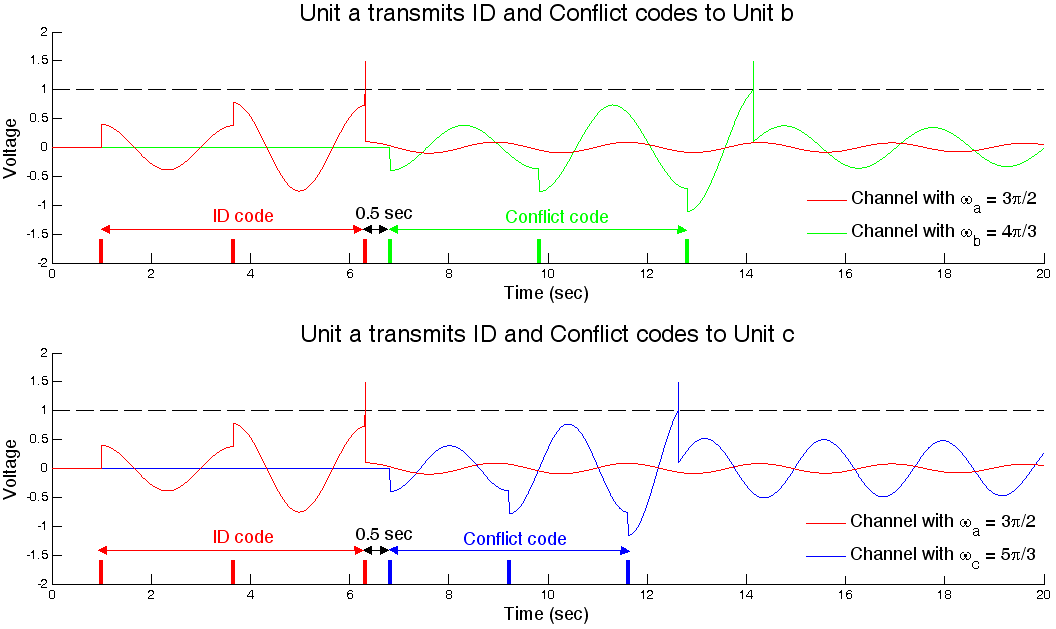}
\caption{Activity of unit $a$.}
\label{unitA}       
\end{figure*}

\begin{figure*}
\includegraphics[height=7cm]{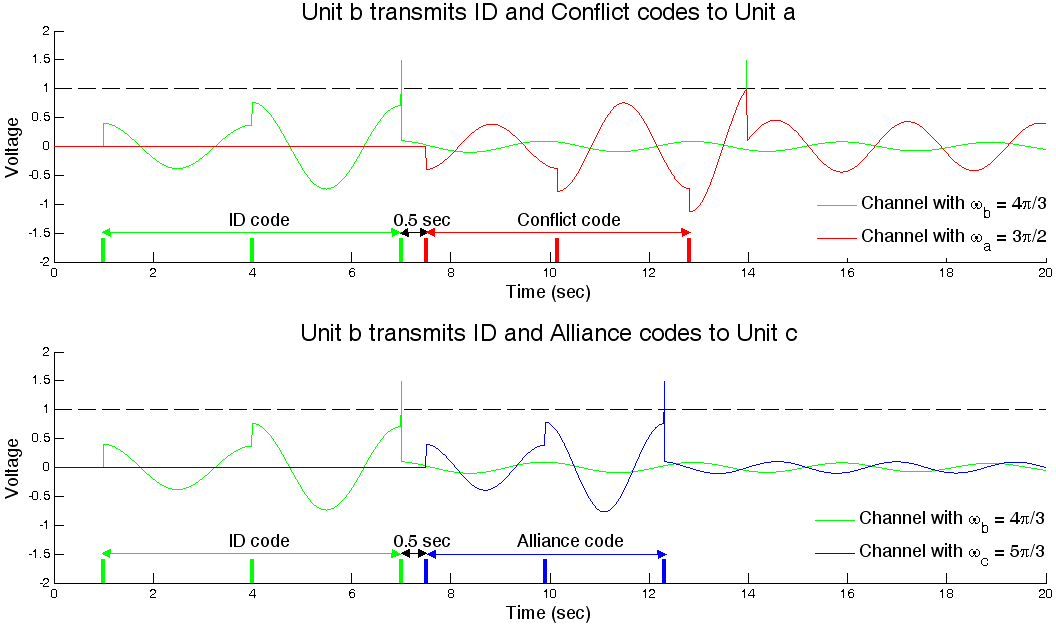}
\caption{Activity of unit $b$.}
\label{unitB}       
\end{figure*}

\begin{figure*}
\includegraphics[height=7cm]{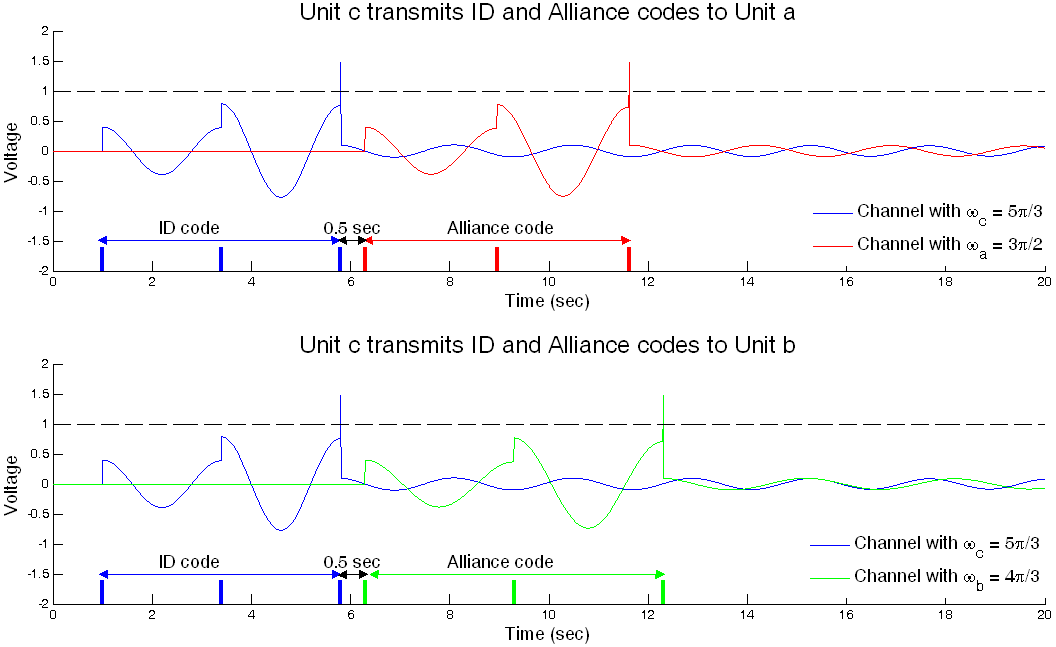}
\caption{Activity of unit $c$}
\label{unitC}       
\end{figure*}

According to Table \ref{relIns}, unit $a$ will send conflict code to both units $b$ and $c$ (Fig.~\ref{unitA}). Unit  $b$ will send conflict code to unit $a$ and alliance code to unit $c$ (Fig.~\ref{unitB}). Unit $c$ will send alliance code to both units $b$ and $c$ (Fig.~\ref{unitC}).

After the codes have been transmitted from each unit to each unit, the information from Table \ref{relIns} become available to each unit in the group. Using the correlation between conflict and alliance codes and logic 1 and 0, we can rewrite Table \ref{relIns}.

\begin{table}
\caption{Transmitted relationship codes}
\begin{center}
\label{relInst}
\begin{tabular}{|c|c|c|c|}
\hline
{}&a&b&c\\
\hline
\rule{0pt}{12pt}a& - &$0$& 0\\[2pt]
\hline
\rule{0pt}{12pt}b&0&-&1\\[2pt]
\hline
\rule{0pt}{12pt}c&1&1&-\\[2pt]
\hline
\end{tabular}
\end{center}
\label{table:tab1}
\end{table}

Therefore, using informaiton from Table \ref{relInst} together with conjunction and disjunction functions, we obtain the relationships installed between the units: units $b$ and $c$ are in alliance, while unit $a$ is in conflict with both units $b$ and $c$.

Since, the informaiton about the relationships between units is now known by each unit in the group , we can construct the relationship graph (Fig. ~\ref{relGrp}) of the Reflexive Game Theory and obtain the polynomial corresponding to this graph, which is $a+bc$.

\subsection{How to Transfer Information about the Influences}
\label{influence}

In this section, we illustrate how to transmit influences of unit on each other. We use the same approach described in the previous section. The only difference is that instead of the alliance or conflict codes, unit transmits some code associated with a particular alternative.

\textit{Example 2.} Suggest, we have Boolean algebra of four alternatives: $1 = \{\alpha,\beta\}$, $\{\alpha\}$, $\{\beta\}$ and $0 = \{\}$. We arbitrary assign a certain code to each alternative: 

1)  code \{0.2, 0.3, 0.7\} corresponds to alternative $1 = \{\alpha,\beta\}$; 

2) code \{0.7, 0.3, 0.2\} corresponds to alternative $0=\{\}$; 

3) code \{0.5, 0.2, 0.5\} corresponds to alternative $\{\alpha\}$; and 

4) code \{0.3, 0.6, 0.3\} corresponds to alternative $\{\beta\}$. 

To make reference easier, we refer to each code as `alterntive name'-code:  

code \{0.7, 0.3, 0.2\} is called $unit$-$code$; 

code \{0.2, 0.3, 0.7\} is $zero$-$code$; 

code \{0.5, 0.2, 0.5\} is referred as $\{\alpha\}$-$code$; and 

code \{0.3, 0.6, 0.3\} is $\{\beta\}$-$code$.

\begin{figure*}
\includegraphics[height=7cm]{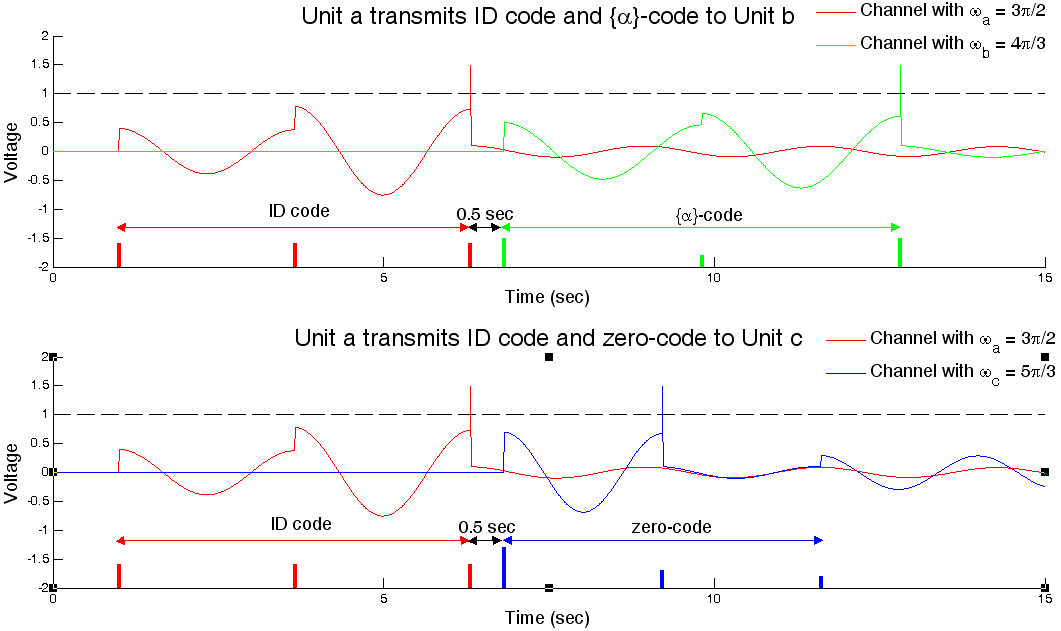}
\caption{Transmission of influnces produced by unit $a$.}
\label{infA}       
\end{figure*}

\begin{figure*}
\includegraphics[height=7cm]{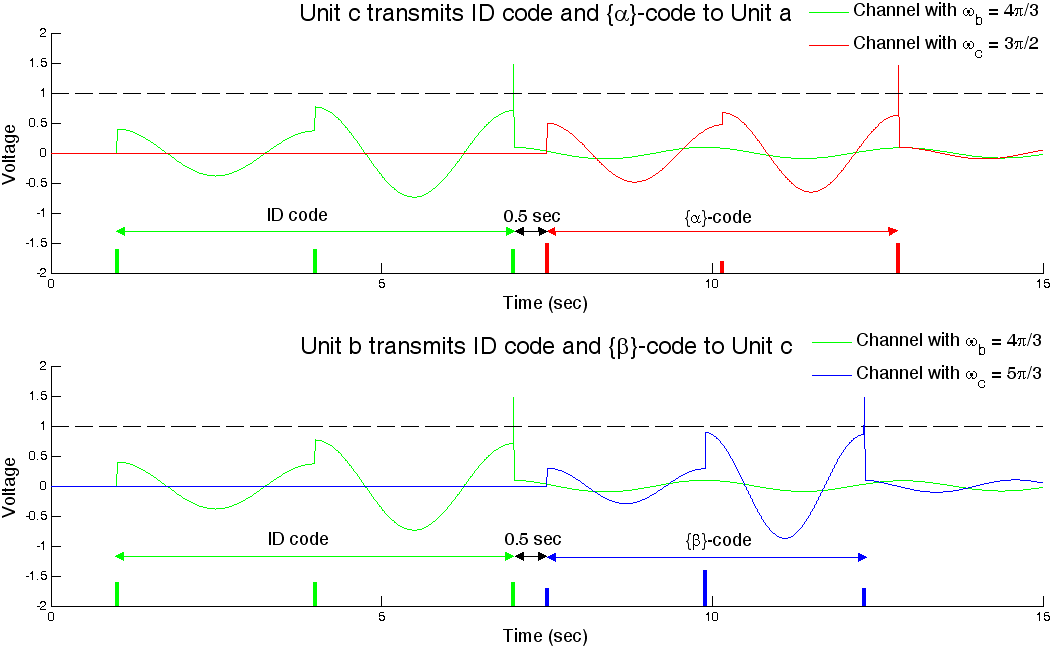}
\caption{Transmission of influnces produced by unit $b$.}
\label{infB}       
\end{figure*}

\begin{figure*}
\includegraphics[height=7cm]{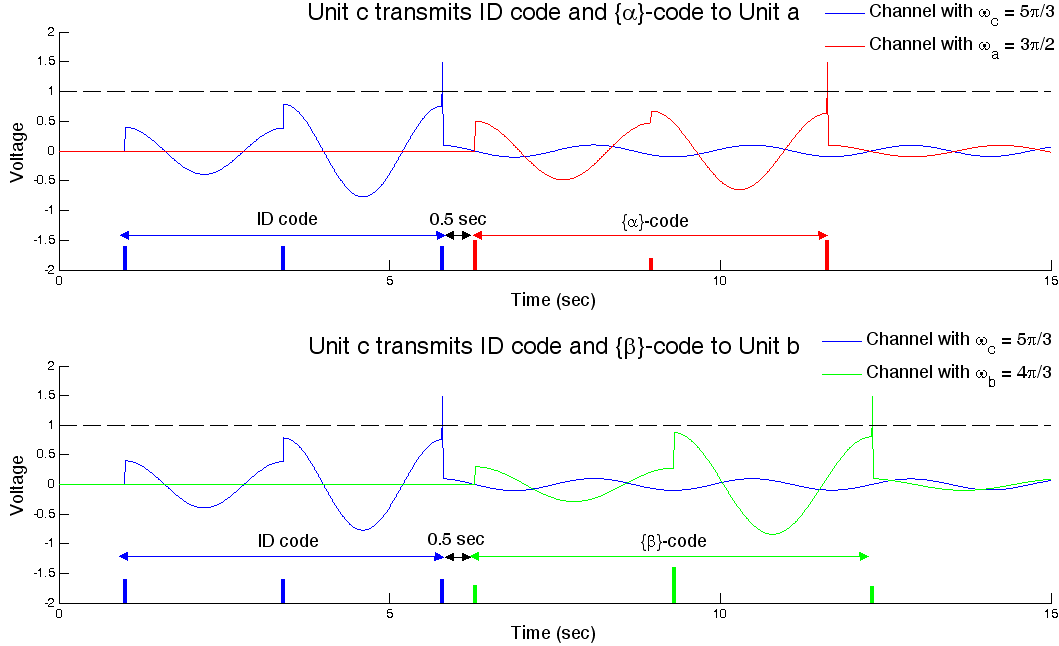}
\caption{Transmission of influnces produced by unit $a$.}
\label{infC}       
\end{figure*}

We assume that unit $a$ makes influences $\{\alpha\}$ and $0=\{\}$ on units $b$ and $c$, respectively (Fig.~\ref{infA}).
Unit $b$ makes influence $\{\alpha\}$ on both units $a$ and $c$ (Fig.~\ref{infB}). Unit $c$ makes influences $\{\alpha\}$ and $\{\beta\}$ on units $a$ and $b$, respectively (Fig.~\ref{infC}).

\begin{table}
\caption{Influence matrix}
\begin{center}
\label{infMat}
\begin{tabular}{|c|c|c|c|}
\hline
{}&a&b&c\\
\hline
\rule{0pt}{12pt}a& a &$\{\alpha\}$& $\{\}$\\[2pt]
\hline
\rule{0pt}{12pt}b&$\{\alpha\}$&b&$\{\beta\}$\\[2pt]
\hline
\rule{0pt}{12pt}c&$\{\beta\}$&$\{\}$&c\\[2pt]
\hline
\end{tabular}
\end{center}
\label{table:tab1}
\end{table}

\subsection{RGT Inference}
Therefore, after all the influences have been transmitted, we obtain the influence matrix (Table \ref{infMat}).

Thus, each unit now has complete information to apply the RGT inference schema based on the decision equations \cite{taras}.
The canonical form of decision equation for unit $a$ is $a = a + bc\overline{a}$ and the corresponding solution interval is $1 \supseteq a \supseteq bc$. The canonical form of decision equation for unit $b$ is $b = (a+c)b + a\overline{b}$ and the corresponding solution interval is $(a+c) \supseteq b \supseteq a$. The canonical form of decision equation for unit $c$ is $c = (a+b)c + a\overline{c}$ and the corresponding solution interval is $(a+b) \supseteq c \supseteq a$.

Under the given influences, the choice of unit $a$ is define by the interval $1 \supseteq a \supseteq \{\}$. The solution interval for unit $b$ turns into equality $ b = \{\alpha\}$. The choice of unit $a$ is define by the interval $\{\beta\} \supseteq a \supseteq \{\}$.

\section{Cooperative Behavoir of the Autonomous Units Controlled by the RGT Algorithms}
\label{cooper}
Until now, we have briefly described the gist of the RGT, communication system for the autonomous units and illustrated how the units can use the information obtained by means of the communication system for the RGT inference. 

In this section we consider how units can make mutual influences in order to achieve a particular goal in a cooperative behavior task.

\textit{Example 3.} Consider four autonomous units (robots). Let these robots are functioning by using the electric accumulators. There's a chaging pool in the restricted perimeter (castle), which has only one entrance (gate). There are three robots $a$, $b$ and $c$ in the castle, and one robot $d$ outside the perimeter.

Let robots $a$, $b$ and $c$ are in alliance with each other and in conflict with robot $d$. Robots $a$, $b$ and $c$ are locked inside the perimeter, but they can open the gate if each robot agrees so. The power source has a limited capacity is only 25\% full.

The accumulators of robots $a$, $b$ and $c$ are 50\% full, while accumulator of robot $d$ is only 10\% full. The power source should be regenerated. However, the time left until regeneration exceed the life-time of the robot $d$'s accumulator.

Consider `open the gates' is action 1 (unit-code). Then Boolean algebra of alternatives contains two elements 1 - to open the gates, and 0 - `don't open'.

From the point of view of Game Theory, it is clearly out of utility to share the power source with `exhausted' enemy robot, but let's have look what should happen if RGT inference is applied.

\begin{figure}
\centering
\includegraphics[height=2cm]{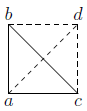}
\caption{Relationship graph for Example 3.}
\label{fig4}
\end{figure}

The diagonal form transformation is present as follows:
\[\begin{array}{*{20}{c}}
   {} & {} & {[a][b][c]} & {} & {} &  \\
   {} & {[abc]} & {} & { + [d]} & {}  \\
   {[abc + d]} & {} & {} & {} & { = abc + d}   \\
\end{array}\]

The resultant decision equations in canoncal forms are:
\begin{eqnarray}
a =(bc+d)a + d\overline{a}  \\
b =(ad+c)b + d\overline{b}  \\
c =(ab+d)c + d\overline{c}  \\
d =d + abc\overline{d} 
\end{eqnarray}

All three robot inside the perimeter are willing not to open the gates and they inlfuence by zero-code on each other and on robot $d$. Then the deicision intervals for each robot are
\begin{eqnarray*}
(bc+d) \supseteq a \supseteq d \Rightarrow  (0+d) \supseteq a \supseteq d \Rightarrow a = d; \\
(ac+d) \supseteq b \supseteq d \Rightarrow  (0+d) \supseteq b \supseteq d \Rightarrow b = d;  \\
(ab+d) \supseteq c \supseteq d \Rightarrow  (0+d) \supseteq b \supseteq d \Rightarrow c = d;  \\
1 \supseteq d \supseteq abc \Rightarrow  1 \supseteq d \supseteq 0.
\end{eqnarray*}

Therefore, the decision of all three robot inside the perimeter is defined by the influence of robot $d$. Therefore, it robot $d$ makes influence 1, then all three robot will agree to open the gate, and since this is required condition, they will open the gate. On the other hand, robot $d$ has \textit{a freedom of choice}.

Therefore, if robot $d$ asks for help, other three robots should open the gate and allow access to the power supply source.

\section{Discussion}
\label{diss}

In this study, we have presented the structure of autonomous units, which allows these units to install communication with each other and create groups. As the basis for communication network, we use resonate-and-fire neurons, which are employed as signal receivers. The main feature of resonate-and-fire neurons is their selectivity to a particular frequency, which is eigen-frequency of the neuron. Therefore, it is possible to send different codes through the same network and be sure that each unit understands the message addressed exclusisvely to it. I do not discuss here physical mechanisms of generating signals.

We illustrated how it is possible to arrage a group of three units as a communication network. I also showed how to code different messages such as sender identification and Boolean algebra alternatives.

We concluded with examples of how a simple group can be arranged based on the information about relationships between units and showed how to transmit the information about influences in the group. Thus, having received the information about the structure of the group and the mutual influences, each autonomous unit can apply algorithms of RGT inferences. Thus, each unit can make both its own choice and also predict the possible choices of other members of the group.  Therefore, the fusion of the proposed sketch of communication network with the RGT inference allows to obviously demostrate principles of the RGT on the particular autonomous units. \\

\end{document}